\documentclass[prl,reprint,twocolumn,showpacs,preprintnumbers,amsmath,amssymb, nofootinbib,floatfix]{revtex4-1}
\usepackage[T1]{fontenc}
\usepackage[latin9]{inputenc}
\usepackage{float}
\usepackage{graphicx}
\usepackage{esint}
\usepackage{dcolumn}
\usepackage[tight]{subfigure}
\usepackage{verbatim}
\usepackage{xspace}
\usepackage{mathtools}
\usepackage{etoolbox}
\usepackage{epstopdf}
\usepackage{epsfig}
\usepackage{amsmath}
\usepackage{amssymb}
\usepackage{amsfonts}
\usepackage{mathptmx}
\usepackage{eucal}
\usepackage{bm}
\usepackage{color}
\usepackage{eucal}
\usepackage{color}
\usepackage[colorlinks=true,linktoc=all,linkcolor=blue,breaklinks=true, citecolor=blue,urlcolor=blue]{hyperref}
\usepackage{natbib}
\usepackage{siunitx}

\begin{document}

\title{Superconducting Quantum Interference Single-Electron Transistor}

\author{Emanuele Enrico$^{1}$}
\email{e.enrico@inrim.it}
\author{Francesco Giazotto$^{2}$}
\email{francesco.giazotto@sns.it}
\affiliation{$^{1}$ INRIM, Istituto Nazionale di Ricerca Metrologica, Strada delle Cacce
91, I-10135 Torino, Italy}
\affiliation{$^{2}$NEST, Istituto Nanoscienze-CNR and Scuola Normale Superiore, Piazza
S. Silvestro 12, Pisa I-56127, Italy}

\begin{abstract}
We propose the concept of a quantized single-electron source based on the interplay between Coulomb blockade and magnetic flux-controllable superconducting proximity effect.
We show that flux dependence of the induced energy gap in the density of states of a nanosized metallic  wire can be exploited as an efficient tunable energy barrier which enables charge pumping configurations with enhanced functionalities. 
This control parameter strongly affects the charging landscape of a normal metal island with non-negligible Coulombic energy. 
Under a suitable evolution of a time-dependent magnetic flux the structure behaves likewise a turnstile for single electrons in a fully electrostatic regime.
\end{abstract}

\maketitle
Synchronized transport of charge quanta has been envisaged since the very beginning of \emph{single electronics} \cite{AverinLikharev1986}, i.e., circuits where manipulation of single electrons can be performed.
So far a number of quantum effects have  been exploited in solid-state devices to obtain a fine control over electromagnetic quantities in view of the realization of their quantum standards \cite{Flowers1324}.
This technology has been exploited in a wide range of applications, covering on-chip cooling \cite{PhysRevLett.98.037201,PhysRevLett.99.027203,PhysRevLett.103.120801}, single photons detection \cite{app6020035} and current sources \cite{PekolaRevModPhys2013}.
The performance of single electron current sources is a trade-off between current amplitude and its accuracy \cite{PekolaRevModPhys2013,Giblin2012}, and from the high sensitivity of these structures to both background charge fluctuations \cite{Zimmerman2014} and residual microwave radiation in the cryogenic setup \cite{PhysRevLett.105.026803}.
Yet, different approaches have been conceived to overcome these limitations.  Most of them relies on the so-called Coulomb-blockade effect which can be tuned by a locally-applied electric field through capacitively-coupled gates \cite{Pekola2008,PhysRevB.77.153301,PekolaRevModPhys2013,Blumenthal2007,Giblin2012}, whereas only few schemes are based on a hybrid electric- and magnetic-driven clocking \cite{Niskanen2003,PhysRevB.86.060502}. In this latter context it is also worth mentioning two fully magnetic-field-driven concepts, i.e., a ferromagnetic single-electron pump \cite{PhysRevB.64.235418}, and a Josephson quantum electron pump \cite{Giazotto2011}.

One recent promising proposal is based on the interplay between the superconducting energy gap and the charging energy in hybrid single-electron transistors (HSETs) \cite{PekolaRevModPhys2013,Pekola2008}.
The turnstile operation of such a device originates from a time-periodic voltage applied to a gate electrode capacitively-coupled to a small metallic island, the preferential tunneling direction through the structure being guaranteed by a finite source-drain bias
voltage. 
A different approach has been applied to two-dimensional electron gas-based charge pumps. In this context, gate electrodes create the junctions barriers which are shaped in a time-dependent fashion, allowing transport of a single electron per cycle by taking advantage of the Coulombic energy of the island \cite{Giblin2012}.
These charge pumps operate in a zero-bias configuration, the directionality of events being controlled by properly shaping in time the barriers.

\begin{figure}[ht!]
\begin{centering}
\includegraphics[width=\columnwidth]{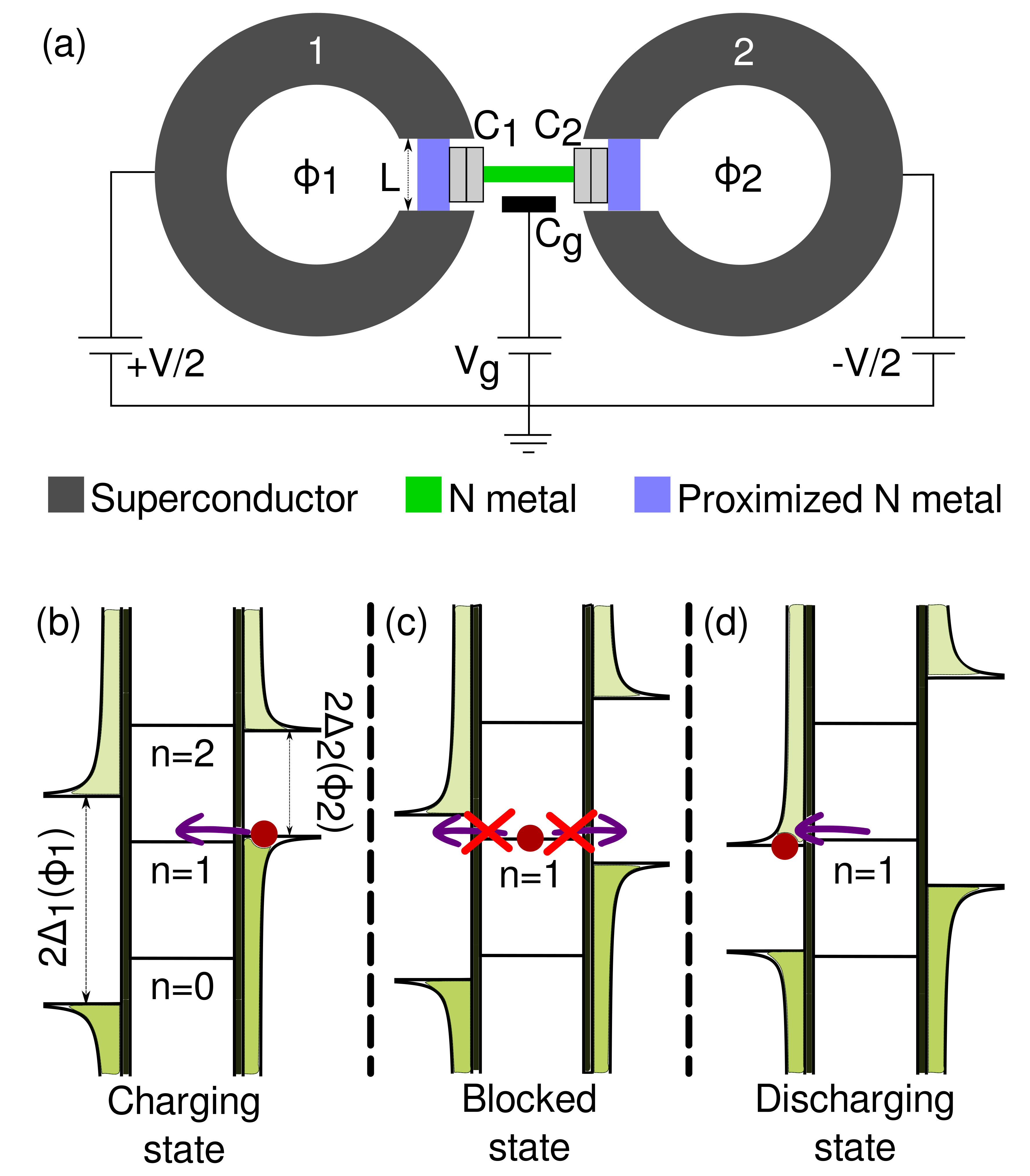}
\vspace{-2mm}
\par\end{centering}
\caption{
(a) Schematic of a  SQUISET. 
Left (1) and right (2) superconducting leads act as source and drain, respectively,  whereas the central gate electrode is capacitively-coupled to the N metal island. 
The latter is connected to the lateral proximity N regions via tunnel junctions.
The loops are threaded by $\varPhi_{i}$ ($i=1,2$) magnetic fluxes. 
(b-d) Sketches of the low-temperature energy band diagrams of the SQUISET biased at $eV=\Delta_{0}$ under a sweep of magnetic flux with  $\varPhi_{1}-\varPhi_{2}=\varPhi_{0}/2$.
$\Delta_0$ is the zero-temperature superconducting energy gap, and
 $\Phi_0$ is the flux quantum. 
$\varPhi_{1}$ was set to 0.15$\varPhi_{0}$ (b),  0.25$\varPhi_{0}$(c) and 0.35$\varPhi_{0}$(d).
The island charging energy ($E_{\mathrm{c}}$) leads to a discrete level spacing vertically shifted by the gate electrode ($n_{\mathrm{g}}=$ \SI{0.5}).
For the above schemes we set $E_{\mathrm{c}}=\Delta_{0}$. 
\label{fig:Schematics}}
\end{figure}

On the basis of the above described strategies, and exploiting recent advances  in magnetic flux-tunable proximity effect as an effective building block to implement  phase coherent superconductor-normal  metal (SN) structures \cite{Giazotto2010,DAmbrosio2015,PhysRevApplied.2.024005,PhysRevB.84.214502}, we put  forward the concept of a quantized single-electron turnstile where the flux-dependent proximity gap created in a N nanowire acts as a tunable barrier coupled to a Coulomb-blockaded island. In such a structure the magnetic flux can drive an opaque NIS junction to an NIN one, where I denotes an insulator. 
Under this premise,  this single-charge structure will be referred to as superconducting quantum interference single-electron transistor (SQUISET).

We investigate a simple design for the SQUISET implementation by following Fig. \ref{fig:Schematics}(a).
In particular, we assume the structure to be symmetrical and composed by two identical superconduting quantum interference proximity transistors (SQUIPTs) \cite{Giazotto2010,DAmbrosio2015} pierced by different magnetic fluxes $\varPhi_{1}$ and $\varPhi_{2}$, which play the role of source ($1$) and drain ($2$) electrodes. 
Furthermore,  the SQUIPTs are connected by a normal metal (N) island through two identical tunnel junctions of capacitance $C$ and resistance $R_{\mathrm{T}}$. 
The island is capacitively-coupled to a gate electrode at voltage $V_{\mathrm{g}}$ via the capacitance $C_{\mathrm{g}}$ which induces $n_{\mathrm{g}}=C_{\mathrm{g}}V_{\mathrm{g}}/e$ elementary charges on it. 
The structure is symmetrically biased with a voltage $V$, and we suppose the charging energy ($E_c$) of the island to be dominated by the capacitance of the junctions, $E_{\mathrm{c}}=e^{2}/2C_{\Sigma}$ where $C_{\Sigma}=2C+C_{\mathrm{g}}\approx2C$.
Each SQUIPT is composed by a superconducting loop interrupted by a diffusive N wire of length $L$, and negligible transverse dimensions \cite{NotaNW1}. 
In the following we assume the wire as quasi-one-dimensional, and its contacts with the S ring as perfectly-transmitting interfaces. 
Moreover, we consider the case of a \emph{short} N bridge satisfying the condition $E_{\mathrm{th}}\gg\Delta_0$, where $E_{\mathrm{th}}=\hbar D/L^2$ is the Thouless energy, $D$ is the wire diffusion constant, and $\Delta_0$ is the zero-temperature order parameter of the S loops.
In such a regime an analytic expression for the wire density of states (DoS) can be derived therefore simplifying the transistor analysis.
In addition, the SQUISET performance is optimized in this limit since proximity effect in the N wires is maximized. 
In this case the DoS $\nu_i$ ($i=1,2$) in each N proximized region is given by \cite{Heikkila2002}
\begin{equation}
\nu_{i}(E,\Phi_{i})=\left|\text{Re}\left[\frac{E+i\Gamma_{i}}{\sqrt{ \left(E+i\Gamma_{i}\right)^{2}-\Delta_{i}^{2}(T)\mathrm{co}\mathrm{s}^{2}(\pi\varPhi_{i}/\varPhi_{0}) } }\right]\right|,\label{eq:1-DOS}
\end{equation}
where $\varGamma_{i}$ (set to $10^{-5}\Delta_{0}$ in all the calculations) models the inelastic scattering rate under the relaxation time approximation \cite{Dynes1984}, and $\Phi_0$ is the flux quantum.
From Eq. \ref{eq:1-DOS} it follows that the DoS in the N wires shows a BCS-like shape with a flux-dependent induced gap, $\Delta_{i,g}(T,\varPhi_{i})=\Delta_{i}(T)\left|\mathrm{cos}\left(\pi\varPhi_{i}/\varPhi_{0}\right)\right|$.
In particular, for $\Phi_i=\Phi_0/2$ the gap is fully closed.

Insight into the operation principle of the SQUISET can be gained by looking at Fig. \ref{fig:Schematics}(b-d) which shows a sketch of the device clocking cycle via three energy diagrams corresponding to different magnetic flux values. 
The discrete energy levels generated by $E_c$ are depicted in the island whereas  flux-controllable energy gaps $\Delta_i(\Phi_i)$ are represented in both source and drain electrodes. 
By changing the energy distribution of free and occupied states, the magnetic flux can open or close independently tunneling channels in the left and right junction, the bias voltage  imposing directionality to single-electron current. 

The flux-dependent tunneling rates between the leads and the island can be evaluated within the ``orthodox theory'' of single-electron tunneling \cite{AverinLikharev1986} as
\begin{equation}
\begin{array}{c}
\varGamma_{1,n}^{\pm}\left(\varPhi_{1}\right)=\frac{1}{e^{2}R_{T}}\int\mathrm{d}E\nu_{1}\left(E,\varPhi_{1}\right)f(E)\left[1-f(E-E_{1,n}^{\pm})\right]\\
\\
\varGamma_{2,n}^{\pm}\left(\varPhi_{2}\right)=\frac{1}{e^{2}R_{T}}\int\mathrm{d}E\nu_{2}\left(E,\varPhi_{2}\right)f(E+E_{2,n}^{\pm})\left[1-f(E)\right],
\end{array}\label{eq:TunnelingRates}
\end{equation}
naming $E_{i,n}^{\pm}=\pm2E_{\mathrm{c}}(n-n_{\mathrm{g}}\pm1/2)\pm eV/2$ the free energy variation as a consequence of tunneling events through the $i$th junction which increase ($+$) or decrease ($-$) the number of excess charges on the island.
In Eqs. (\ref{eq:TunnelingRates}) we assume no energy exchange with the environment \cite{GrabertDevoret1992}, and we consider both the leads and the island to be in equilibrium at temperature $T$.
\begin{figure}
\begin{centering}
\includegraphics[width=1\columnwidth]{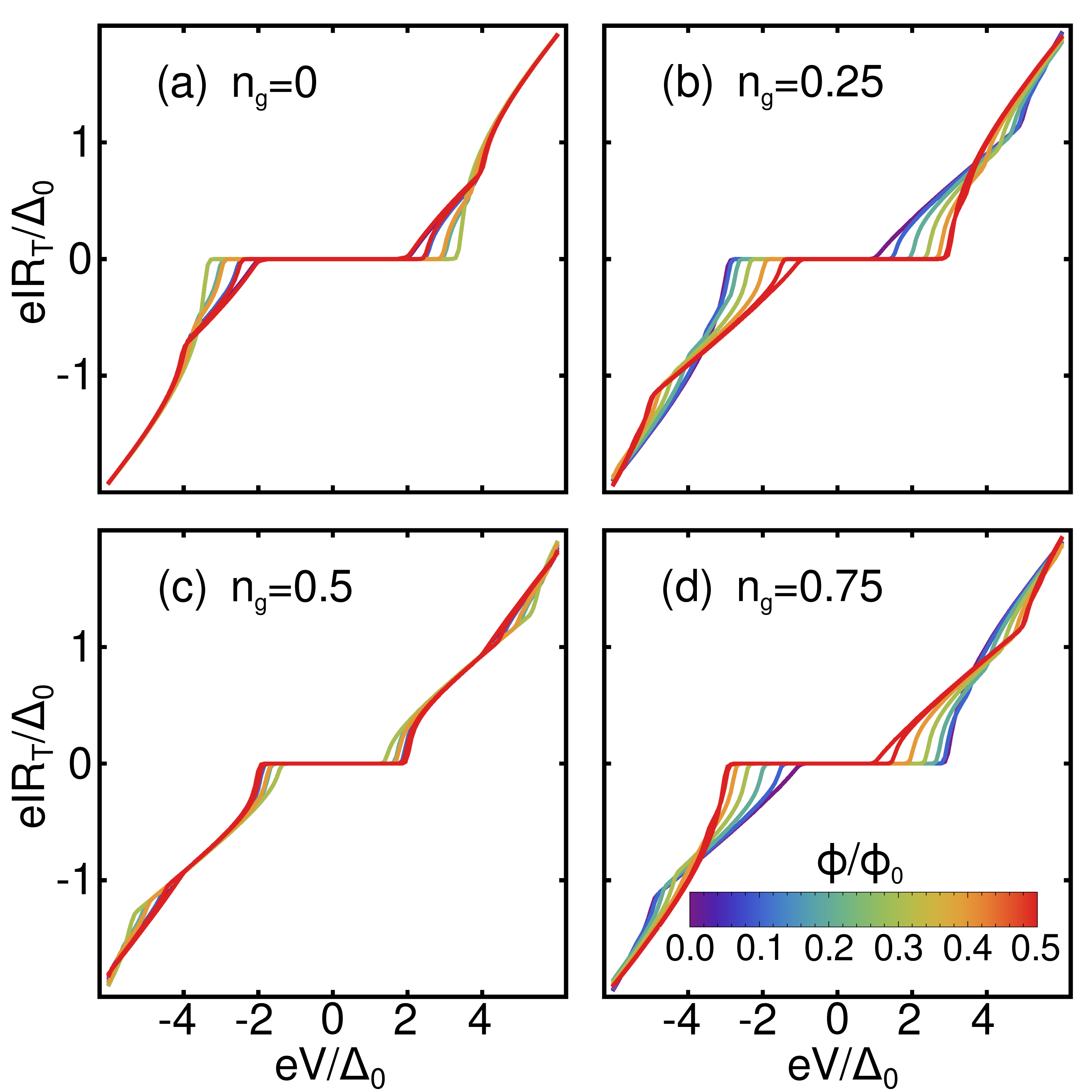}\vspace{-2mm}
\par\end{centering}
\caption{
(a-d) DC current vs voltage ($IV$) characteristics of a SQUISET calculated at different magnetic fluxes and for different $n_{g}$ values. $\varPhi_{1}$ and $\varPhi_{2}$ are supposed to be related by $\varPhi_{1}-\varPhi_{2}=\varPhi_{0}/2$.
Aluminum (Al)  is the superconductor chosen for the loops, and we set $\Delta_{1}\left(0\right)=\Delta_{2}\left(0\right)=\Delta_{0}=E_{\mathrm{c}}=$\SI{185}{\micro\electronvolt}.
The turnstile operates at thermal equilibrium at \SI{35}{\milli\kelvin}.
The electric current is blockaded by the two energy gaps in a plateau shifted and modulated in amplitude by the magnetic flux.\label{fig:IVcurves}}
\end{figure}

\begin{figure*}[t!]
\begin{centering}
\includegraphics[width=\textwidth]{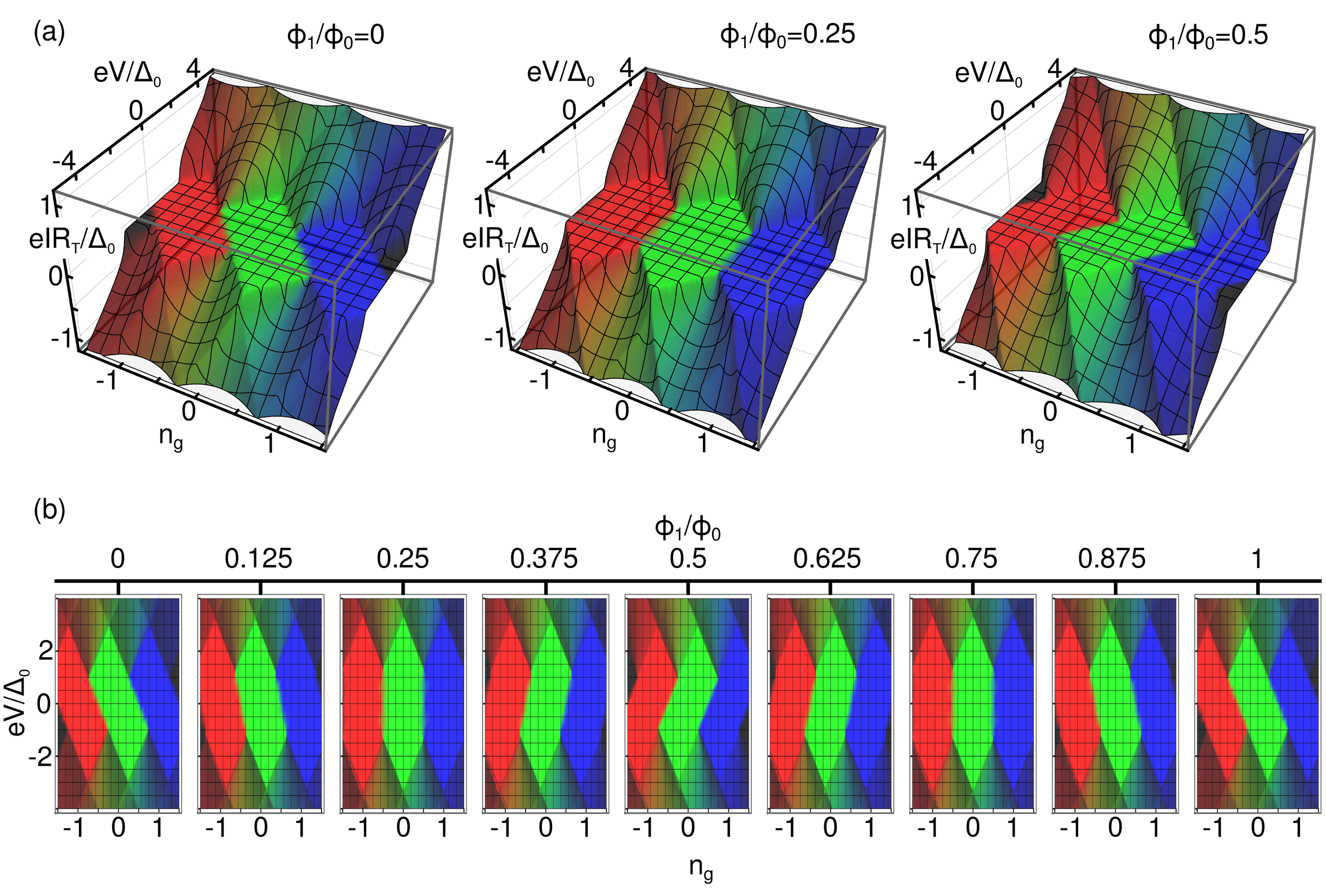}\vspace{-2mm}
\par\end{centering}
\caption{(a) Stability diagrams at different $\varPhi_{1}/\varPhi_{0}$ values
(\SI{0}, \SI{0.25} and \SI{0.5} from left to right) in a SQUISET with $\varPhi_{1}-\varPhi_{2}=\varPhi_{0}/2$.  The plots surfaces have been colored having the red,
green and blue channels proportional to $p_{-1}$, $p_{0}$ and $p_{1}$,
respectively. (b) Top view of the stability diagrams in a complete $\varPhi_{1}/\varPhi_{0}$ flux period.  Here we set $\Delta_{1}\left(0\right)=\Delta_{2}\left(0\right)=\Delta_{0}=E_{\mathrm{c}}=50k_{\mathrm{B}}T=$
\SI{185}{\micro\electronvolt}. \label{fig:StabilityDiagrams}}
\end{figure*}

In order to study deterministically the dynamics of the system we focus on the sequential tunneling regime. In this framework, following a Fokker-Planck approach for a particular time-dependent flux driving signal, we can write the master equation in the matrix form for the probability $p_{n}(t)$ to store $n$ charges in excess on the island as a function of time
\begin{equation}
\frac{\mathrm{d}p_{n}}{\mathrm{d}t}=\sum_{m}\varGamma_{nm}p_{m},\label{eq:MasterEquation}
\end{equation}
being
\begin{equation}
\begin{array}{c}
\varGamma_{nm}=\delta_{m,n-1}\left[\varGamma_{1,m}^{+}+\varGamma_{2,m}^{+}\right]+\delta_{m,n+1}\left[\varGamma_{1,m}^{-}+\varGamma_{2,m}^{-}\right]-\\
-\delta_{m,n}\left[\varGamma_{1,m}^{+}+\varGamma_{1,m}^{-}+\varGamma_{2,m}^{+}+\varGamma_{2,m}^{-}\right]
\end{array}\label{eq:GammaNM}
\end{equation}
the tunneling rate between $n$ and $m$ state.  
In its stationary version, the master equation can be used to evaluate source-drain charge current as a function of static control parameters $\varPhi_{i}$, $V$ and $n_{\mathrm{g}}$ through the relation $I=-e\sum_{n}p_{n}\left(\varGamma_{1,n}^{+}-\varGamma_{1,n}^{-}\right)$.
The family of curves displayed in Fig. \ref{fig:IVcurves} represents the static calculation performed for selected values of $n_{\mathrm{g}}$ and different magnetic flux $\varPhi_{i}$. 
We set $\varPhi_{1}=\varPhi_{2}+\varPhi_{0}/2$, and the difference $\varPhi_{0}/2$ may arise either from geometrical construction of the SQUIPT loops or from an applied  non-uniform static magnetic field. 

For low enough temperature ($k_{\mathrm{B}}T\ll E_{\mathrm{c}}$), the current is blockaded up to a voltage resulting from the threshold relations $E_{i,n}^{\pm}>-\Delta_{i,g}(T,\varPhi_{i})$ which come from the energy gap induced on the $i$th junction. 
The imposed flux asymmetry yields
\begin{equation}
\begin{array}{c}
E_{1,n}^{\pm}>-\Delta_{1,g}(0)\left|\mathrm{cos}\left(\pi\varPhi_{1}/\varPhi_{0}\right)\right|,\\
\\
E_{2,n}^{\pm}>-\Delta_{2,g}(0)\left|\mathrm{sin}\left(\pi\varPhi_{1}/\varPhi_{0}\right)\right|.
\end{array}\label{eq:Threshold}
\end{equation}
These equations illustrate a crucial point: the magnetic flux affects only the energy thresholds, and therefore assumes the role of external control parameter. 
In Fig. \ref{fig:IVcurves}(a)  ($n_{\mathrm{g}}=0$), the free-energy variations for a single electron tunneling event are $E_{i,n}^{\pm}=\pm2E_{\mathrm{c}}(n\pm1/2)\pm eV/2$. By considering the lowest energy contribution coming from $n=0$, the voltage thresholds become $\left|eV\right|<2E_{\mathrm{c}}+2\mathrm{Min}\left[\Delta_{1,g}(0)\left|\mathrm{cos}\left(\pi\varPhi_{1}/\varPhi_{0}\right)\right|,\Delta_{2,g}(0)\left|\mathrm{sin}\left(\pi\varPhi_{1}/\varPhi_{0}\right)\right|\right]$.
In the case of Fig. \ref{fig:IVcurves}(c) ($n_{\mathrm{g}}=0.5$), by taking into account that both $n=0$ and $n=1$ are energetically possible for different magnetic flux values, leads to the opposite situation where the current is blocked for
$\left|eV\right|<2\mathrm{Max}\left[\Delta_{1,g}(0)\left|\mathrm{cos}\left(\pi\varPhi_{1}/\varPhi_{0}\right)\right|,\Delta_{2,g}(0)\left|\mathrm{sin}\left(\pi\varPhi_{1}/\varPhi_{0}\right)\right|\right]$.
In the last condition the charging energy is analytically canceled by the effect of the gate which positions the device in a regime where the current is blockaded (for each $\varPhi_{1}$) by the stronger of the two induced gaps. 
Therefore the two junctions can be driven alternatively and independently in open or closed states, and the island consequently in $n=0$ or $n=1$, by the sole operation of the magnetic flux. 

Figures \ref{fig:IVcurves}(b,d) better illustrate the decoupling action of the two gaps caused by the charging energy. 
For $n_{\mathrm{g}}=0.25$ ($n_{\mathrm{g}}=0.75$) the decoupling is maximum so that two different thresholds for positive voltage $0<eV<E_{\mathrm{c}}+2\Delta_{1,g}(0)\left|\mathrm{cos}\left(\pi\varPhi_{1}/\varPhi_{0}\right)\right|$ ($0<eV<E_{\mathrm{c}}+2\Delta_{2,g}(0)\left|\mathrm{sin}\left(\pi\varPhi_{1}/\varPhi_{0}\right)\right|$), and for negative voltage $-E_{\mathrm{c}}-2\Delta_{2,g}(0)\left|\mathrm{sin}\left(\pi\varPhi_{1}/\varPhi_{0}\right)\right|<eV<0$ ($-E_{\mathrm{c}}-2\Delta_{1,g}(0)\left|\mathrm{cos}\left(\pi\varPhi_{1}/\varPhi_{0}\right)\right|<eV<0$) can be identified with the same procedure. These latter considerations essentially lead to an almost rigid voltage shift of the blockaded region represented in Fig. \ref{fig:IVcurves}(b,d).

Equations (\ref{eq:Threshold}) are usually represented in a three-dimensional stability diagram showing the electric current vs $V$ and $n_{\mathrm{g}}$. 
Although for the SQUISET the control parameter ($\varPhi_{1}$) would require a further dimension for the stability diagram to be fully illustrated, 
one can easily follow this additional dependence in Fig. \ref{fig:StabilityDiagrams} where we have selected three representative conditions. 
Here the current surfaces have been colored imposing the red, blue and green color channels proportional to $p_{-1}$, $p_{0}$ and $p_{1}$, respectively.
It clearly appears how not just the boundaries are tuned by the flux but also the blockade ``diamonds'' are deformed, and in some way ``rotated'', in the $V$-$n_{\mathrm{g}}$ space. 
The three diagrams are essentially showing the SQUISET configuration that starts from NINIS-like state (with $\varPhi_{1}=$ \SI{0}), passes through a SINIS-like (at $\varPhi_{1}/\varPhi_{0}=$ \SI{0.25}), eventually reaching a SININ-like behavior ($\varPhi_{1}/\varPhi_{0}=$ \SI{0.5}).
While in the regions of current interdiction the island charging configuration ($n$) is by definition fixed to almost unitary values (colored then by a single RGB channel color), as soon as a finite current starts to flow the island experiences a time sequence of different $n$-states resulting
in a intermediate color.

\begin{figure}
\begin{centering}
\includegraphics[width=1\columnwidth]{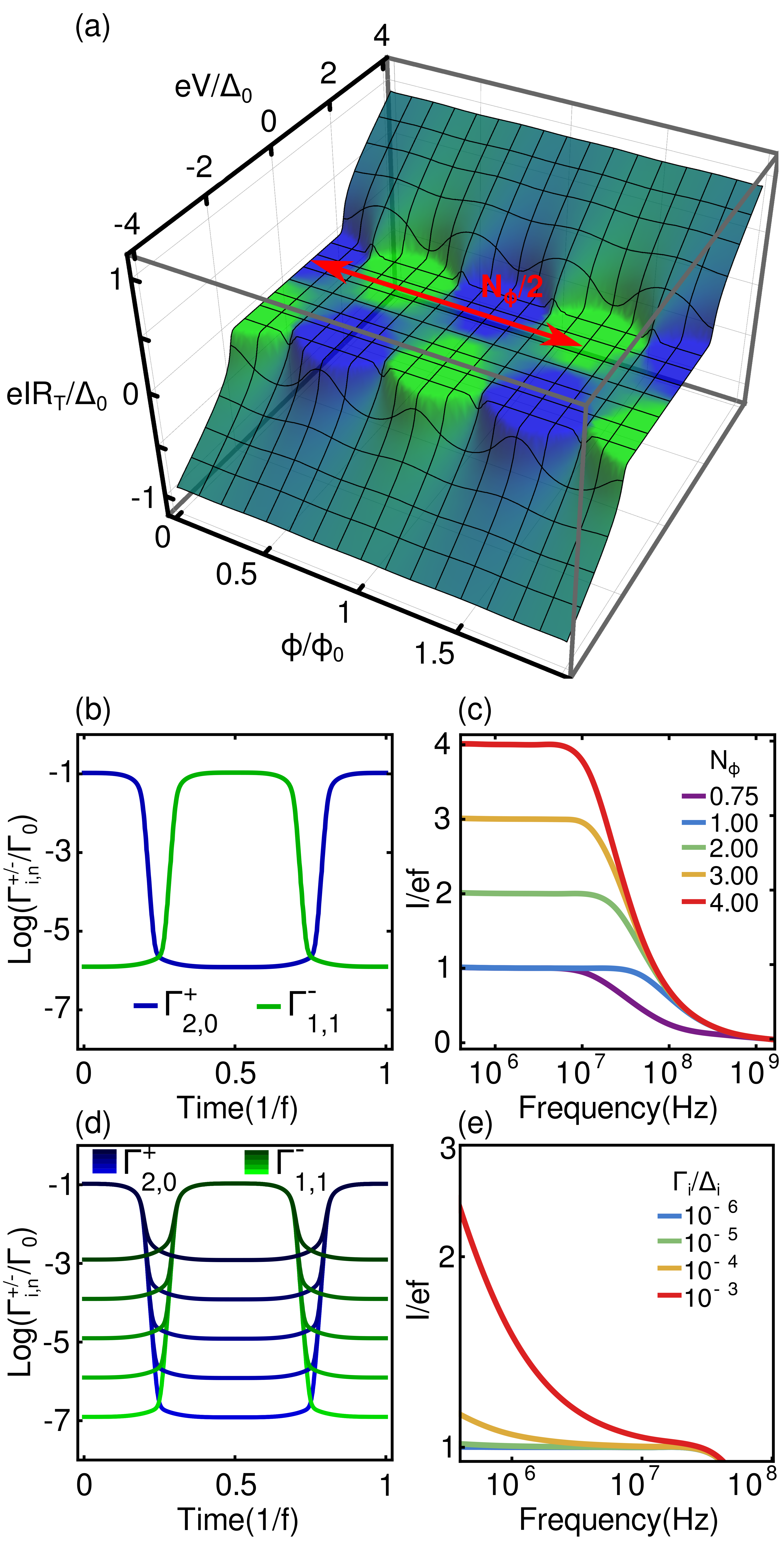}\vspace{-2mm}
\par\end{centering}
\caption{
(a) SQUISET stability diagram vs $\varPhi=\varPhi_1=\varPhi_2+\varPhi_0/2$ and $V$ at fixed gate voltage ($n_{\mathrm{g}}=$ \SI{0.5}).
The surface has been colored having the green and blue channels proportional to $p_{0}$ and $p_{1}$, respectively, and shows the periodicity in the blockade regime states as a function of magnetic flux. 
The red arrow represents Eq. (\ref{eq:Fluxes}) with $N_{\varPhi}=3$. 
(b) Basic pumping cycle of a SQUISET with $N_{\varPhi}=$ \SI{1} in a fully symmetrical configuration having $\Delta_{1}\left(0\right)=\Delta_{2}\left(0\right)=\Delta_{0}=E_{\mathrm{c}}=$ \SI{185}{\micro\electronvolt}. Here $\varGamma_{2,0}^{+}$ and $\varGamma_{1,1}^{-}$, normalized by $\varGamma_{0}\equiv\Delta_{0}/e^{2}R_{\mathrm{T}}$, represent the dominant rates moving the system between $p_{0}$ and $p_{1}$ states.
During this particular cycle from Eq. (\ref{eq:Fluxes}) the bias voltage is set to $V=\Delta_{0}/e$, and $n_{\mathrm{g}}=0.5$. 
(c) Frequency dependence of the normalized source-drain current for different $N_{\varPhi}$ values. (d)  Basic pumping cycle of a SQUISET as in (b) with different values of the Dynes parameters $\log{(\varGamma_{i}/\Delta_{i})}$, respectively $-7$,$-6$,$-5$,$-4$,$-3$ from bottom (lighter) to top (darker) curves. (e) Normalized source-drain current for $N_{\varPhi}=1$ at different values of the   Dynes parameters  $\varGamma_{i}/\Delta_{i}$. \label{fig:StabilityDiagram-Frequency}}
\end{figure}

As mentioned before, under particular circumstances ($n_{\mathrm{g}}=$\SI{0.5}), the flux parameter can drive the SQUISET into a charging state [Fig. \ref{fig:Schematics}(b)] or into a discharging state [Fig. \ref{fig:Schematics}(d)] passing through a blocked state [Fig. \ref{fig:Schematics}(c)]. 
In this situation the control parameters are then reduced to $\varPhi_{1}$ and $V$ only, 
and a new kind of stability diagram can be introduced [see Fig. \ref{fig:StabilityDiagram-Frequency}(a)]. 
Here, the green channel is proportional to $p_{0}$ and the blue one to $p_{1}$ so that it is clear how the device can settle to different stable regions having the bias threshold $\left|eV\right|<2\mathrm{Max}\left[\Delta_{1,g}(0)\left|\mathrm{cos}\left(\pi\varPhi_{1}/\varPhi_{0}\right)\right|,\Delta_{2,g}(0)\left|\mathrm{sin}\left(\pi\varPhi_{1}/\varPhi_{0}\right)\right|\right]$
introduced before with a flux periodicity of $1/2\varPhi_{0}$. 
Figure \ref{fig:StabilityDiagram-Frequency}(a) shows how a closed trajectory in the $\varPhi_{1}-V$ space, and resulting in a single-electron net current per cycle, can be found.

Apart the above stationary approximation, Eq. (\ref{eq:GammaNM}) allows to calculate [see Fig. \ref{fig:StabilityDiagram-Frequency}(b)] the time evolution of the rates affecting the state of the island during a particular magnetic flux cycle defined, for instance,  as 
\begin{equation}
\varPhi_{1}\left(t\right)=\varPhi_{2}\left(t\right)+\frac{\varPhi_{0}}{2}=-\frac{\varPhi_{0}}{4}\cdot N_{\varPhi}\cdot\left[\mathrm{tri}\left(f\cdot t\right)-1\right],
\label{eq:Fluxes}
\end{equation}
where $\mathrm{tri}(x)$ is the triangular waveform function. 
$N_{\varPhi}$ denotes the peak-to-peak amplitude of the time-dependent modulation in units of $\varPhi_{0}/2$ [see the red arrow in Fig. \ref{fig:StabilityDiagram-Frequency}(a)]. 
Equation (\ref{eq:Fluxes}) is only one of the possible clocking cycles, and it has been chosen here for the sake of clarity. 
It drives the induced gaps in source and drain electrodes from a complete closing to the situation where the N wires fully inherit a superconducting behaviour from the S loops.
Several flux sequences transferring one charge per period can be easily found. 
We note that even time-dependent cycles yielding an uncomplete closing of the induced gaps can drive efficiently the single-charge clocking mechanism under suitable biasing conditions.
From Eq. (\ref{eq:Fluxes}) the rates are then essentially alternating for the two NIS junctions  [see Fig. \ref{fig:StabilityDiagram-Frequency}(b)]; moreover, they are almost constant in ``closed'' states and dominated by the intra-gap leakage of the induced gaps, while in the ``open''
state they reach a nearly unitary value in unit of $\varGamma_{0}\equiv\Delta_{0}/e^{2}R_{\mathrm{T}}$.
The origin the leakage stems from the smeared DoS [see Eq. (\ref{eq:1-DOS})] modeling the environmental-assisted tunneling.

Following the system evolution by solving Eq. (\ref{eq:MasterEquation}) we have calculated the probabilities $p_{n}(t)$ during this particular control parameter cycle as a function of frequency for $eV=\Delta_{0}$ and $n_{\mathrm{g}}=0.5$, and starting from a reasonable initial condition $p_{n}(0)=\delta_{n,0}$. 
After few cycles the system reaches a periodic quasi-equilibrium in which the occupation probabilities clearly oscillate from $p_{1}$ to $p_{0}$ state. 
In full analogy with the turnstile behavior of an HSET driven by a radio-frequency gate voltage, the SQUISET oscillates between $p_{1}\sim$ \SI{1} state to $p_{0}\sim$ \SI{1} state, leaving spurious clocking proportional only to exponentially-suppressed tunneling events.
This particular cycle is depicted in Fig. \ref{fig:Schematics}(b-d) where in Fig. \ref{fig:Schematics}(b) the full branch of the drain DoS reaches the $n=$ \SI{1} energy level leading to a ``charging state'' where a single quasiparticle can tunnel into the island. 
In the intermediate period, shown in Fig. \ref{fig:Schematics}(c), both junctions are in a blockaded state due to the interplay between the charing energy and the energy gaps.
In Fig. \ref{fig:Schematics}(d) the upper, empty branch of the source's DoS aligns to the occupied island state opening the possibility for a ``discharging'' condition where a single electron escapes from the island driving it to the initial $n=$ \SI{0} state. 
In this way one electron per cycle is moved from drain to source generating a net current equal to $\left\langle I\right\rangle =ef$. 

Considering an arbitrary flux modulation amplitude $N_{\varPhi}$, the average source-drain current can be obtained by integrating over one control parameter cycle as follows,
\begin{equation}
\left\langle I\right\rangle =-\frac{e}{T}\int_{0}^{T}\mathrm{d}t\sum_{n}p_{n}\left(t\right)\left(\varGamma_{1,n}^{+}\left(t\right)-\varGamma_{1,n}^{-}\left(t\right)\right).
\label{eq:AverageCurrent}
\end{equation}
Figure \ref{fig:StabilityDiagram-Frequency}(c) shows clear current plateau  in unit of $ef$ for a wide range of flux frequencies up to a cutoff which is inversely proportional to $N_{\varPhi}$ in the case of integer values of $N_{\varPhi}$. 
The SQUISET acts thereby in a single-electron turnstile fashion moving $\left[N_{\varPhi}\right]$ charges each cycle,  being $\left[N_{\varPhi}\right]$ the $N_{\varPhi}$ nearest integer. 
The maximum generated current is limited by the $R_{\mathrm{T}}C$ time constant associated to a single electron tunneling event essentially responsible for the missed tunneling errors. 
This proportionality holds for the triangular waveform suggested in Eq. (\ref{eq:Fluxes}) which maximizes the cutoff frequency, superimposing an instantaneous current across each junction with a $\left(f\cdot[N_{\varPhi}]\right)^{-1}$ periodicity in the
time domain. 
In this view, under the driving expressed by Eq. (\ref{eq:Fluxes}) the SQUISET realizes the relation $\left\langle I\right\rangle =\left[N_{\varPhi}\right]\cdot ef$.

Figure \ref{fig:StabilityDiagram-Frequency}(d) clearly shows the impact of different Dynes parameters on the dominant rates in the turnstile configuration. At higher values the ratio between the unwanted rate and the clocking rate increases proportionally to the Dynes parameter itself leading to a leakage current as shown in Fig. \ref{fig:StabilityDiagram-Frequency}(e). As in the case of the SINIS turnstile, the accuracy of our SQUISET device increases by improving the ideality of the superconducting leads DoS. 

For the SQUISET implementation we exploited the peculiar behavior of Eq. (\ref{eq:1-DOS}) making the NIS junctions in a HSET-like structure tunable by introducing an additional control parameter, the magnetic flux. 
The pure quantum nature of flux-tunable phase interference in a proximized nanowire guides the single-electron tunneling in the semiclassical regime. 
In this view our SQUISET adds new perspectives to metallic single electronics introducing different clocking configurations which can be of interest as building blocks in fields other than quantized current generation, like coherent caloritronics \cite{RevModPhys.78.217,Giazotto2012bis,MartinezPerez2015quatris,QuarantaAPL,GiazottoHeatModulatorAPL,MartinezPerezAPL} or quantum information technology \cite{Nielsen:2011:QCQ:1972505}.
Further investigation on pumping accuracy is crucial including higher order tunneling processes for which the static condition $n_{\mathrm{g}}=$ \SI{0.5} could potentially limit unwanted events to two orders of magnitude lower than in the SINIS turnstile \cite{PhysRevLett.101.066801}. 
Eventually, the coupling with the environmental residual radiation as a source of intra-gap leakage \cite{PhysRevLett.105.026803} and the background charge fluctuation sensitivity of the SQUISET as a fully electro-statical single-electron clocking device \cite{PhysRevB.53.13682} are still unexplored phenomena in the field of phase-coherent single electronics.

We are pleased to thank G. Amato and L. Callegaro for useful comments. E.E. acknowledges partial financial support from the European Metrology
Research Programme (\textquotedblleft EXL03 MICROPHOTON\textquotedblright ). The EMRP is jointly funded by the EMRP participating countries within EURAMET and the European Union.
F.G. acknowledges  the European Research Council under the European Union's Seventh Framework Programme (FP7/2007-2013)/ERC grant agreement No. 615187-COMANCHE for partial financial support.

\bibliographystyle{apsrev4-1}

\begin{thebibliography}{34}
\bibitem [{Not()}]{NotaNW1}%
  \BibitemOpen
  \href@noop {} {}\bibinfo {note} {The nanowire could be in principle also of S
  type with $L\lesssim \xi_{0}$, where $\xi_{0}$ is the superconducting
  coherence length}\BibitemShut {NoStop}%
\end{thebibliography}

%

\end{document}